\documentclass[pre,article]{revtex4}%
\usepackage{graphicx}
\usepackage{amsmath}
\usepackage{amsfonts}
\usepackage{amssymb}%
\providecommand{\U}[1]{\protect\rule{.1in}{.1in}}
\providecommand{\U}[1]{\protect\rule{.1in}{.1in}}
\begin{document}
\title{\ Neutrino oscillations as many-particle \\induced interference between distinguishable particles}
\author{A. Cabo$^{1}$, N. Cabo Bizet$^{2}$,}
\affiliation{$^{1}$\textit{Theoretical Physics Department, \\Instituto de Cibern\'{e}tica,
Matem\'{a}tica y F\'{\i}sica, \\Calle E, No. 309, Vedado, La Habana, Cuba. \\}}
\affiliation{$^{2}$\textit{Departamento de F\'{\i}sica, DCI,\\ Campus Leon, Universidad de
Guanjuato, \\ CP. 37150, Leon, Guanajuato.\\ }}
\begin{abstract}
\noindent\ We investigate the causes of the curious property of neutrino
oscillations of looking as an interference between fully distinguishable
particles. The sources of this effect are identified as to be determined by the
many particle nature of space of states of the quantum field theory. It is
firstly underlined that in order to explain the observed oscillations, the
neutrino subspace of states should not be interpreted as the direct product of
the three neutrinos Fock spaces, which is equivalent to imposing a
superselection rule. Further, it is argued that the linear completion of such
a direct product space of states permits to describe the measured
oscillations. Thus,  the space of states of the Standard Model (SM)
should be interpreted as the linear completion of the direct product of all he Fock spaces (boson or
fermions ones) associated to the distinguishable particles. Then, it follows
that the neutrino oscillations are determined as interference between
independent particles, which are generated by states being outside the direct
product of  the neutrino Fock spaces. This interpretation seems to imply a
large amount of interference observations between  distinguishable particles in
Particle as well as in Condensed Matter Physics. For seeking clearness, the
discussion is done in the framework of a simple Quantum Field Theory (QFT)
model of two relativistic
free massive neutrinos.

\end{abstract}
\maketitle

\section{Introduction}

The interference effect between  distinguishable particles had been observed long time ago
and started to be investigated in the original works \cite{piccioni,Gellman,Cabbibo,pontecorvo}.
 The particular case of the   neutrino oscillations  has been and continues to be a relevant theme
of research in Particle Physics. Since its prediction by B. Pontecorvo in
references \cite{pontecorvo,pontecorvo1,pontecorvo2}, the effect had been
theoretically as well as experimentally intensively investigated. The
discovery of the real occurrence in nature of these oscillations, in super
Kamiokande and Solar neutrino observations was a breakthrough step
\cite{kamiokande, solar}. After it, an enormous amount of investigations about
this effect had been performed
\cite{Anka1,Anka2,GiuntiKim,BilenkyGiunti,Giunti1,
Giunti2,Giunti3,Grimus,Beuthe,AkhmedovKopp,BlasoneVitiello,BlasoneGargiulo,GiuntiFock,AkhmedovSubtetlies}%
. The oscillations had been studied either through Quantum Mechanics (QM) and
Quantum Field Theory (QFT) methods
\cite{BlasoneVitiello,GiuntiFock,AkhmedovKopp,Beuthe}. In our view the QFT
methods had contributed to clarify some of the assumptions which have been  done in the
QM approaches. In particular the question about the possibility of defining a
Fock space for the flavor eigenfunctions for the electron, muon and tau
neutrinos as expressed as linear superpositions of the really stable
propagating neutrino models, had been extensively discussed \cite{GiuntiFock}.
This aspect constitutes an example of a question related with neutrino
oscillation physics that today remain under discussions
\cite{AkhmedovSubtetlies}. One point which in our view deserves research
attention is related with the fact that the neutrino oscillations look as a
surprising interference effect, similar to the one occurring in the QM of a single particle,
but occurring between fully distinguishable fermion particles, each one of
them described by a wavefunction being  in a separate Fock space \cite{pontecorvo}.
This is a peculiar effect if we consider the idea often used in QM
presentations about that different (distinguishable) particles do not
interfere between them. Thus, it seem of interest to identify the reason why
the interference between neutrinos oscillations discards this frequent
consideration adopted in QM.

In this work we address this question. For this purpose the related issue
about the proper interpretation of the space of states in quantum field theory
is examined. In particular we search for the possibility of describing the
quantum oscillations involving different distinguishably particles in the many
particle space of states. A model involving only two distinguishable types
(flavors) of relativistic particles is considered for the sake of clearness.
In first place, it is argued that to describe interference between two fermion
flavors in analogy with the neutrino oscillations measurements, the space of
states should not be defined as the direct product of the two Fock spaces
associated to the two fermion particles. That procedure will be equivalent to
impose a superselection rule for defining the allowed physical states. This is
because to employing the direct product as the space of states leads to accept
the nonphysical character of the state obtained after acting with the addition
of two field operators (each one creating a particle the two separate Fock
spaces) over the vacuum state. The explanation for this exclusion is the fact
that the states generated by the action of such additions do not pertain to
the direct product of the two Fock spaces.

Afterwards, it is considered that the space of the states of the model is the
linear completion of the direct product of the two Fock states. In other
words, the linear combination with arbitrary coefficients of the external
products of the states in each of the two Fock spaces. It should be stressed
that the use of this space of states is naturally suggested, if we do not
assume the presence of any superselection rule to restrict the superposition
principle of the quantum field theory. It is then verified that this space
allows to define the addition of field operators acting in each separate Fock
space, as generating new and physically allowed states. Further, rotated
flavor creation and annihilation operators are defined by linear combinations
of the creation and annihilation operators for stable propagating neutrino
modes solving the Dirac equation. This procedure allows to argue that states
generated by these rotated flavor fields at a given time, are linear
combinations of the propagating particles modes. Moreover, it follows that
under a measurement in such a propagating states the probability of its
measurement oscillates. Note again, that after accepting the completed direct
product as the space of states, the linear superposition of field operators
acting in different Fock spaces are valid entities in terms of which Hermitian
physical quantities can be properly defined,

Therefore, the presented discussion makes clear that the usual representation
of QFT allows to predict quantum oscillations between various distinguishable
particles, explaining in this way, how the many particle quantum theory
defined by the QFT properly explains the curious interference between
different particles. Henceforth, the analysis suggests the validity of what we
consider a relevant property of QFT generalization of the Quantum Mechanics:
the opening of the possibility for observing interference between almost all
kinds of distinguishable particles. This property comes from the fact that the
superposition principle in QFT, allows to add fields being defined in
different Fock spaces (describing distinguishable particles) and then the
existence of interference effects determined by superposition of the many
particles space of states should lead to many types of observable oscillations.

The plan of the presentation is as follows. In Section 2, the two relativistic
neutrino free QFT is presented. The Hamiltonian expressed in terms of the two
fields, and the commutation properties among these are written. This allow to
define the creation and annihilation operators in momentum space for each the
two flavors and their commutation relations. Section 3, defines two possible
the space of states of the QFT associated to the model. The direct product of
the two Fock states associated to each of the particles and its linear
completion. Finally, Section 4 shows how the linear completion space of states
is able to describe the oscillation between the two types neutrinos in the
considered model. The Summary reviews the discussion and results.

\section{The non relativistic two neutrino model}

Let us consider the mentioned in the Introduction model of two free
relativistic massive fermions with flavor indices $\nu=1,2$. We will start
form  the quantum theory defined by the Hamiltonian operator
\begin{equation}
H=\sum_{\nu=1,2}\int d\mathbf{x}\text{ }\overline{\psi}_{\nu}(\mathbf{x,}%
t)(-i\mathbf{\gamma.\nabla}+m)\psi_{\nu}(\mathbf{x,}t),
\end{equation}
expressed in terms of \ a four components $\ r=1,2,3,4$ fermion field
$\psi(\mathbf{x,}t)$%
\begin{equation}
\psi(\mathbf{x,}t)=\psi_{r}(\mathbf{x,}t)\equiv\left(
\begin{tabular}
[c]{l}%
$\psi_{1}(\mathbf{x,}t)$\\
$\psi_{2}(\mathbf{x,}t)$\\
$\psi_{3}(\mathbf{x,}t)$\\
$\psi_{4}(\mathbf{x,}t)$%
\end{tabular}
\ \ \ \right)  .
\end{equation}

The fields in terms of the annihilation and creation operators $b_{_{\nu,s}%
}(\mathbf{p}),b_{_{\nu,s}}(\mathbf{p}),$ for the two types of particles
$\nu=1,2$, having helicities $s=\pm1,$ and the corresponding annihilation and
creation operators for their antiparticles $d_{_{\nu,s}}(\mathbf{p}%
),d_{_{\nu,s}}^{+}(\mathbf{p})$,$\nu=1,2,$  have the usual expansions
\begin{align}
\psi_{\nu}(\mathbf{x,}t)  &  =\sum_{\mathbf{p,}\nu,s}(\omega_{\nu
,s}(\mathbf{p},\mathbf{x})\text{ }b_{_{\nu,s}}(\mathbf{p})+v_{\nu,s}%
^{r}(\mathbf{p},\mathbf{x})\text{ }d_{_{\nu,s}}^{+}(\mathbf{p})),\\
\psi_{\nu}^{+}(\mathbf{x,}t)  &  =\sum_{\mathbf{p,}\nu,s}(\omega_{\nu,s}%
^{\ast}(\mathbf{p},\mathbf{x})\text{ }b_{_{\nu,r,s}}^{+}(\mathbf{p})+v_{\nu
,s}^{\ast}(\mathbf{p},\mathbf{x})\text{ }d_{_{\nu,s}}(\mathbf{p})),
\end{align}
in which the $\omega_{\nu,s}(\mathbf{p})$ are the positive energy solutions of
the Dirac equation with helicity $s=\pm1$ for each of the two flavors
$\nu=1,2,$ defined as \cite{schweber}
\begin{align}
\ w_{\nu,+}(\mathbf{p,x})  &  =\frac{\exp(i\text{ }\mathbf{p.x})}{\sqrt{L^{3}%
}\sqrt{2}\sqrt{2(n_{3}+1)}}\frac{\sqrt{\epsilon_{\nu}(\mathbf{p})+m_{\nu}}%
}{\epsilon_{\nu}(\mathbf{p})}\left(
\begin{tabular}
[c]{l}%
$n_{3}+1$\\
$n_{1}+in_{2}$\\
$\frac{|\mathbf{p|}}{\epsilon_{\nu}(\mathbf{p})+m_{\nu}}\left(
\begin{tabular}
[c]{l}%
$n_{3}+1$\\
$n_{1}+in_{2}$%
\end{tabular}
\ \right)  $%
\end{tabular}
\ \right)  ,\text{ \ }\\
\text{\ }w_{\nu,-}(\mathbf{p,x})  &  =\frac{\exp(i\text{ }\mathbf{p.x})}%
{\sqrt{L^{3}}\sqrt{2}\sqrt{2(n_{3}+1)}}\frac{\sqrt{\epsilon_{\nu}%
(\mathbf{p})+m_{\nu}}}{\epsilon_{\nu}(\mathbf{p})}\left(
\begin{tabular}
[c]{l}%
$-n_{1}+in_{2}$\\
$n_{3}+1$\\
$\frac{|\mathbf{p|}}{\epsilon_{\nu}(\mathbf{p})+m_{\nu}}\left(
\begin{tabular}
[c]{l}%
$-n_{1}+in_{2}$\\
$n_{3}+1$%
\end{tabular}
\ \right)  $%
\end{tabular}
\ \right)
\end{align}

The antiparticle malfunctions $v_{\nu,+}(\mathbf{p})$ with helicity $s=\pm1$
for each of the two flavors $\nu=1,2,$have \ the expressions%
\begin{align}
\ v_{\nu,+}(\mathbf{p})  &  =\frac{\exp(-i\text{ }\mathbf{p.x})}{\sqrt{L^{3}%
}\sqrt{2}\sqrt{2(n_{3}+1)}}\frac{\sqrt{\epsilon_{\nu}(\mathbf{p})+m_{\nu}}%
}{\epsilon_{\nu}(\mathbf{p})}\left(
\begin{tabular}
[c]{l}%
$-\frac{|\mathbf{p|}}{\epsilon_{\nu}(\mathbf{p})+m_{\nu}}\left(
\begin{tabular}
[c]{l}%
$n_{3}+1$\\
$n_{1}+in_{2}$%
\end{tabular}
\ \ \right)  $\\
$n_{3}+1$\\
$n_{1}+in_{2}$%
\end{tabular}
\ \ \right)  ,\text{ \ }\\
\text{\ }v_{\nu,-}(\mathbf{p})  &  =\frac{\exp(-i\text{ }\mathbf{p.x})}%
{\sqrt{L^{3}}\sqrt{2}\sqrt{2(n_{3}+1)}}\frac{\sqrt{\epsilon_{\nu}%
(\mathbf{p})+m_{\nu}}}{\epsilon_{\nu}(\mathbf{p})}\left(
\begin{tabular}
[c]{l}%
$-\frac{|\mathbf{p|}}{\epsilon_{\nu}(\mathbf{p})+m_{\nu}}\left(
\begin{tabular}
[c]{l}%
$-n_{1}+in_{2}$\\
$n_{3}+1$%
\end{tabular}
\ \ \right)  $\\
$-n_{1}+in_{2}$\\
$n_{3}+1$%
\end{tabular}
\ \ \right)  .
\end{align}

The energies associated to the two types of particles are
\begin{equation}
\epsilon_{\nu}(\mathbf{p})=\sqrt{\mathbf{p}^{2}+m_{\nu}^{2}},\text{ \ }%
\nu=1,2.
\end{equation}

The field operators and the creation and annihilation ones for the two kinds
of particles and antiparticles satisfy
\begin{align}
\left[  \psi_{\nu}(\mathbf{x,}t),\psi_{\nu^{\prime}}^{+}(\mathbf{x}^{\prime
}\mathbf{,}t)\right]  _{+} &  =I\text{ }\delta_{\nu,\nu^{\prime}}%
\delta(\mathbf{x-x}^{\prime}),\\
\left[  b_{\nu,s}(\mathbf{p}),b_{\nu^{\prime},s^{\prime}}^{+}(\mathbf{p}%
^{\prime})\right]  _{+} &  =\delta_{\nu,\nu^{\prime}}\delta_{s,s^{\prime}%
}\delta_{\mathbf{p,p}^{\prime}}^{(K)},\\
\left[  d_{\nu,s}(\mathbf{p}),d_{\nu^{\prime},s^{\prime}}^{+}(\mathbf{p}%
^{\prime})\right]  _{+} &  =\delta_{\nu,\nu^{\prime}}\delta_{s,s^{\prime}%
}\delta_{\mathbf{p,p}^{\prime}}^{(K)},\\
\left[  b_{\nu,s}(\mathbf{p}),b_{\nu^{\prime},s^{\prime}}(\mathbf{p}^{\prime
})\right]  _{+} &  =0,\\
\left[  b_{\nu,s}^{+}(\mathbf{p}),b_{\nu^{\prime},s^{\prime}}^{+}%
(\mathbf{p}^{\prime})\right]  _{+} &  =0,\\
\left[  d_{\nu,s}(\mathbf{p}),d_{\nu^{\prime},s^{\prime}}(\mathbf{p}^{\prime
})\right]  _{+} &  =0,\\
\left[  d_{\nu,s}^{+}(\mathbf{p}),d_{\nu^{\prime},s^{\prime}}^{+}%
(\mathbf{p}^{\prime})\right]  _{+} &  =0.
\end{align}
where, $\delta_{\mathbf{p,p}^{\prime}}^{(K)}$ is the Kronecker Delta%
\begin{equation}
\delta_{\mathbf{p,p}^{\prime}}^{(K)}=%
\begin{tabular}
[c]{l}%
$1\ \ $if\ \ $\ \ \mathbf{p=p}^{\prime}$\\
$0$ \ if \ \ \ \ $\mathbf{p\neq p}^{\prime}$%
\end{tabular}
\ \ \ ,
\end{equation}
and $\mathbf{p}$\textbf{ } are the momenta satisfying periodicity conditions
in a large cubic box having a length size $L$ and volume $L^{3}.$ That is, if
$\ L=Na$ and $N$ is even, the components of $\ $the\ momenta $\ \mathbf{p}%
=(p_{1},p_{2},p_{3})$ are given as%
\begin{align}
p_{1}\text{ } &  =\frac{2\pi}{a}\frac{m_{1}}{N},\text{ }-\frac{N}{2}\leq
m_{1}<\frac{N}{2},\nonumber\\
p_{2}\text{ } &  =\frac{2\pi}{a}\frac{m_{2}}{N},\text{ }-\frac{N}{2}\leq
m_{2}<\frac{N}{2},\nonumber\\
p_{3}\text{ } &  =\frac{2\pi}{a}\frac{m_{3}}{N},\text{ }-\frac{N}{2}\leq
m_{3}<\frac{N}{2}.
\end{align}

\section{The space of states in the quantum field theory}

After constructed the second quantization of the above defined simple non
relativistic two massive neutrinos system, we will consider the main issue in
this work: to investigate the influence of the space of states adopted for the
theory, on the possibility for the description of the interference between
different distinguishable as it occurs between the measured neutrino oscillations.

It can be started by remarking that in the literature, it has been discussed
the possibility that when you have distinguishable particles the correct space
of states of the combined system could be the direct product of the Fock
spaces which is associated to each of the distinguishable particles. In connection with this view, it should be stressed that
the this assumption is equivalent to establish a superselection rule
over states not admitting a physical states, the
addition of states of the different species. The establishment of
superselection rules in QFT is allowed for sure in some cases \cite{schweber}. That is the
situation with respect to the electric charge in which you can adopt to not
allow the superposition of states showing different amount of electric charge.
However, in such cases the exclusion of these type of superpositions is
"dynamically" excluded, since the interaction operators conserve the charge of
the states over which they act. Therefore, if we assume that the initial
states over which the evolution operator acts has a well defined amount of
electric charge, any state after acting over it with evolution operator will
have the same eigenvalue of the charge operator. However, in problems where
the interaction operators have no property restricting the resulting states to
the same physical subspace after their action, it seems not possible to impose
such superselection rules. We had the impression about that this is the
situation in the case of neutrino oscillations, and this idea motivated the
present work.

Then, as it was mentioned, the space of states of the simple QFT model
constructed here will be  examined. The aim  is to determine the
conditions for being able to describe the observed neutrino oscillations.
Below, in the context of the model constructed in the past section, it will be
argued that the appropriate space of states for the QFT description should not
be the direct product of the two Fock spaces defined. In place of it, the sates
should be considered in the linear completion of the direct product of the two
Fock spaces for each of the two distinguishable particles. The conclusion
indicates that the superposition principle for states (the addition of
physical states is a resulting physical state) should not have superselection
rules, in order to describe neutrino oscillations.

Consider the two Fock spaces $\mathcal{F}_{\nu}$ , $\nu=1,2$ generated by the
before defined creation operators for each of the two particles. The states of
a complete basis in the Fock spaces of each separate particles will be
indicated as $\left\vert \Phi_{f_{\nu}}\right\rangle _{\mathcal{F}_{\nu}}%
,\nu=1,2$ where $f_{\nu}$  is an index for any of the states in the Fock space
of type $\nu$ . Then, the states in the direct product of the two Fock spaces
can be written as
\begin{align}
\left\vert \Psi\right\rangle _{\mathcal{F}_{1}\otimes\mathcal{F}_{2}}  &
=\sum_{f_{1}}\sum_{f_{2}}C_{_{f_{1}}}C_{_{f_{2}}}\left\vert \Phi_{f_{1}%
}\right\rangle _{\mathcal{F}_{1}}\otimes\left\vert \Phi_{f_{2}}\right\rangle
_{\mathcal{F}_{2}}\nonumber\\
&  =\left(  \sum_{f_{1}}C_{_{f_{1}}}\left\vert \Phi_{f_{1}}\right\rangle
\right)  _{\mathcal{F}_{1}}\otimes\left(  \sum_{f_{2}}C_{_{f_{2}}}\left\vert
\Phi_{f_{2}}\right\rangle _{\mathcal{F}_{2}}\right)  .
\end{align}
But as mentioned before, this class of states, for $a$ and $b$ different form
zero constants,  excludes  superpositions of the form
\[
a\text{ \ }\left\vert \Phi_{f_{1}}\right\rangle _{\mathcal{F}_{1}}%
\otimes\left\vert \Phi_{f_{2}}\right\rangle _{\mathcal{F}_{2}}+b\text{
\ }\left\vert \Phi_{f_{1}^{\text{ }\prime}}\right\rangle _{\mathcal{F}_{1}%
}\otimes\left\vert \Phi_{f\text{ }_{2}^{\prime}}\right\rangle _{\mathcal{F}%
_{2}},
\]
if ($f_{1},f_{2}$) also differs form ($f_{1}^{\prime},f_{2}^{\prime}$). This
exclusion is related with the fact that the direct product of Fock states is
what is required for to implement a superselection rule. The scalar product of
two states pertaining to the direct product  can be defined as
\begin{equation}
\left\langle \Psi\right.  \left\vert \Psi^{\prime}\right\rangle
_{\mathcal{F}_{1}\otimes\mathcal{F}_{2}}=\sum_{f_{1},}C_{_{f_{1},\text{ }}%
}^{\ast}C_{_{f_{1},\text{ }}}^{\prime}\times\sum_{f_{2}}C_{_{f_{2}\text{ }}%
}^{\ast}C_{_{f_{2}\text{ }}}^{\prime}%
\end{equation}
and normalized states for each component can be defined as separately
satisfying
\begin{align}
\sum_{f_{1},}C_{_{f_{1},\text{ }}}^{\ast}C_{_{f_{1},\text{ }}}  &  =1,\\
\sum_{f_{2}}C_{_{f_{2}\text{ }}}^{\ast}C_{_{f_{2}\text{ }}}  &  =1.
\end{align}

However, there is also the possibility of considering a wider space, the
linear completion of the formerly defined direct product space. This linear
completion, that will be called as $C(\mathcal{F}_{1}\otimes\mathcal{F}_{2})$
can be defined as the set of states generated by the arbitrary coefficients
$C_{_{f_{1},\text{ }f_{2}}}$ in the superposition of the form
\begin{equation}
\left\vert \Psi\right\rangle _{C(\mathcal{F}_{1}\otimes\mathcal{F}_{2})}%
=\sum_{f_{1},f_{2}}C_{_{f_{1},\text{ }f_{2}}}\left\vert \Phi_{f_{1}%
}\right\rangle _{\mathcal{F}_{1}}\otimes\left\vert \Phi_{f_{2}}\right\rangle
_{\mathcal{F}_{2}}.
\end{equation}

It is clear that such states can not be always expressed in the form of a
direct product of linear spaces%
\begin{equation}
\left(  \sum_{f_{1}}C_{_{f_{1}}}\left\vert \Phi_{f_{1}}\right\rangle \right)
_{\mathcal{F}_{1}}\otimes\left(  \sum_{f_{2}}C_{_{f_{2}}}\left\vert
\Phi_{f_{2}}\right\rangle _{\mathcal{F}_{2}}\right).
\end{equation}

The scalar product assumed that the basis states in both Fock spaces are
normalized is defined as%
\begin{equation}
\left\langle \Psi\right.  \left\vert \Psi^{\prime}\right\rangle
_{C(\mathcal{F}_{1}\otimes\mathcal{F}_{2})}=\sum_{f_{1},f_{2}}C_{_{f_{1}%
,\text{ }f_{2}}}^{\ast}C_{_{f_{1},\text{ }f_{2}}}^{\prime}%
\end{equation}
Normalized states satisfy
\begin{equation}
1=\sum_{f_{1},f_{2}}C_{_{f_{1},\text{ }f_{2}}}^{\ast}C_{_{f_{1},\text{ }f_{2}%
}}^{\prime}.
\end{equation}

This definition of the space of states is avoiding  a
superselection rule not allowing the states being expressed as arbitrary linear
combinations of two states in direct product of Fock spaces. Therefore, in general, but in the absence of superselection rules, the Fock space of any QFT of a system of a number of $n_{p}$ distinguishable
particles (being either bosons or fermions) should be interpreted  as the whole set
of states generated by the arbitrary coefficients $C_{_{f_{1},\text{ }%
f_{2},...,..f_{n_{p}}}}$ of the form%
\begin{align}
\left\vert \Psi\right\rangle _{C(\mathcal{F}_{1}\otimes.....\otimes
.\mathcal{F}_{n_{p}})}  &  =\sum_{f_{1},f_{2},...,f_{n_{p}}}C_{_{f_{1},\text{
}f_{2},...,..f_{n_{p}}}}\left\vert \Phi_{f_{1}}\right\rangle _{\mathcal{F}%
_{1}}\otimes\left\vert \Phi_{f_{2}}\right\rangle _{\mathcal{F}_{2}}%
\otimes\text{ }...\text{ }\otimes\left\vert \Phi_{f_{n_{p}}}\right\rangle
_{\mathcal{F}_{n_{p}}}.\\
\left\langle \Psi\right.  \left\vert \Psi^{\prime}\right\rangle
_{C(\mathcal{F}_{1}\otimes.....\otimes.\mathcal{F}_{n_{p}})}  &  =\sum
_{f_{1},f_{2}}C_{_{_{f_{1},\text{ }f_{2},...,..f_{n_{p}}}}}^{\ast}%
C_{_{f_{1},\text{ }f_{2},...,..f_{n_{p}}}}^{\prime}\\
1  &  =\sum_{f_{1},f_{2}}C_{_{_{_{f_{1},\text{ }f_{2},...,..f_{n_{p}}}}}%
}^{\ast}C_{_{_{_{f_{1},\text{ }f_{2},...,..f_{n_{p}}}}}}^{\prime}.
\end{align}

\section{Neutrino oscillations and the space of states}

Let us now consider the QFT defined in section II. The only aspect of the
theory which remains to be defined is the space of the physical states of the
theory. In this case, the many states only including one the two types of
particles (let say of flavor $\nu=1$ or flavor $\nu=2$) are described \ by the
Fock space ($\mathcal{F}_{1}$ or $\mathcal{F}_{2})$ generated by the creation
operators of the specific kind of flavor. Therefore, lets argue below that in
the space of states defined by the linear completion of the direct product of
the two Fock space $C(\mathcal{F}_{1}\otimes\mathcal{F}_{2})$ the neutrino
like oscillations can be effectively described. Conversely, the
oscillations can not be directly explained by assuming the direct product of
the two Fock space as defining the physical space of states for the system

\subsection{Space of states $C(\mathcal{F}_{1}\otimes\mathcal{F}_{2})$}

As they are well defined in this space, we will examine the states of the
form
\begin{align}
\left\vert \Psi\right\rangle _{C(\mathcal{F}_{1}\otimes\mathcal{F}_{2})}  &
=\sum_{s=\pm}C_{_{1,s,}}b_{1,+1}^{+}\text{ \ }\left\vert 0\right\rangle
_{\mathcal{F}_{1}}\otimes\left\vert 0\right\rangle _{\mathcal{F}_{2}%
}+\left\vert \Phi_{f_{1}}\right\rangle _{\mathcal{F}_{1}}\otimes C_{_{f_{2}%
,s}}b_{2,+1}^{+}\left\vert 0\right\rangle _{\mathcal{F}_{2}}\nonumber\\
&  =\sum_{s=\pm}\left(  C_{_{1,s,}}b_{1,+1}^{+}\text{ }(\mathbf{p}%
)+C_{_{f_{2},s}}b_{2,+1}^{+}(\mathbf{p})\right)  \text{\ }\left\vert
0\right\rangle _{\mathcal{F}_{1}}\otimes\left\vert 0\right\rangle
_{\mathcal{F}_{2}},
\end{align}
describing  states in which a one particle state with negative helicity
$s=-1$ is created in the Fock space $\mathcal{F}_{1}$ (with zero particles in
the Fock space $\mathcal{F}_{2}$) is superposed with a zero particle created
in $\mathcal{F}_{1}$ with one particle with helicity $s=-1$ created in
$\mathcal{F}_{2}$ . Both particles have the same momentum $\mathbf{p.}$\ As it
can be noted form the second line of the equation, these states are generated
by a linear combination of field operators describing two different flavor
modes both with a common value of the helicity and momentum. The form of these
states was selected in other to more closely represent the situation for the
neutrino oscillation measurement. The assumption of the physical nature of
these states, then allow to define physical quantities (Hermitian operators
constructed in terms of the employed fields) in terms of these superposition
of fields, which create particles in different Fock spaces as the above
defined ones.

\subsection{Flavor rotated fields}

It is possible to define now flavor rotated fields, as linear functions of the
original fields in terms of which it is possible to define sets of physical
quantities as Hermitian operator constructs. These definitions are here
discussed in order to further consider measurements, describing quantum
oscillations of amplitudes. Let us define the flavor rotated $electron$ and
$muon\ $ like fields $b_{\nu_{e},s}(\mathbf{p})$,$b_{\nu_{\mu},s}%
\ (\mathbf{p})$ \ (which are not the stable neutrino fields $b_{1,s},b_{2,s}$)
as
\begin{align}
b_{\nu_{e},s}\ (\mathbf{p})  &  =\left(  \cos(\theta)b_{1,s}\text{
}(\mathbf{p})+\sin(\theta)b_{2,s}(\mathbf{p})\right)  ,\\
b_{\nu_{e},s}^{+}\ (\mathbf{p})  &  =\left(  \cos(\theta)b_{1,s}^{+}\text{
}(\mathbf{p})+\sin(\theta)b_{2,s}^{+}(\mathbf{p})\right)  ,\\
b_{\nu_{\mu},s}\ (\mathbf{p})  &  =\left(  -\sin(\theta)b_{1,s}\text{
}(\mathbf{p})+\cos(\theta)b_{2,s}(\mathbf{p})\right)  ,\\
b_{\nu_{\mu},s}^{+}\ (\mathbf{p})  &  =\left(  -\sin(\theta)b_{1,s}^{+}\text{
}(\mathbf{p})+\cos(\theta)b_{2,s}(\mathbf{p})\right)  .
\end{align}

These operators, as the previous ones, also satisfy the following commutation
relations
\begin{align}
\left[  b_{\nu_{e},s}(\mathbf{p}),b_{\nu_{e},s^{\prime}}(\mathbf{p}^{\prime
})\right]  _{+}  &  =\delta_{s,s^{\prime}}\delta_{\mathbf{p,p}^{\prime}}%
^{(K)}\\
\left[  b_{\nu_{\mu},s}^{+}(\mathbf{p}),b_{\nu_{\mu},s^{\prime}}%
^{+}(\mathbf{p}^{\prime})\right]  _{+}  &  =0.
\end{align}
We will call $b_{\nu_{e},,s}(\mathbf{p})$ as the electron neutrino field\ of
helicity s and the $b_{\nu_{\mu},,s}(\mathbf{p})$ as the muon neutrino of
helicity $s.$ The inverse transformation takes the form
\begin{align}
b_{1,s}\ (\mathbf{p})  &  =\left(  \cos(\theta)b_{\nu_{e},s}\text{
}(\mathbf{p})-\sin(\theta)b_{\nu_{\mu},s}(\mathbf{p})\right)  ,\\
b_{1,s}^{+}\ (\mathbf{p})  &  =\left(  \cos(\theta)b_{\nu_{e},s}^{+}\text{
}(\mathbf{p})-\sin(\theta)b_{\nu_{\mu},s}^{+}(\mathbf{p})\right)  ,\\
b_{2,s}\ (\mathbf{p})  &  =\left(  \sin(\theta)b_{\nu_{e},,s}\text{
}(\mathbf{p})+\cos(\theta)b_{\nu_{\mu},s}(\mathbf{p})\right)  ,\nonumber\\
b_{2,s}^{+}\ (\mathbf{p})  &  =\left(  \sin(\theta)b_{\nu_{e},s}^{+}\text{
}(\mathbf{p})+\cos(\theta)b_{\nu_{\mu},s}^{+}(\mathbf{p})\right)  .\nonumber
\end{align}

These new operators define creation and annihilation operators of the flavor
rotated state over the vacuum. By example, the creation of a single particle
state with rotated flavor $\nu_{e}$ $\ $or $\nu_{\mu}$ , momentum\textbf{
}$\mathbf{p}$ and helicity $s$ are defined by
\begin{align}
b_{\nu_{e},s}^{+}(\mathbf{p})\left\vert 0\right\rangle _{C(\mathcal{F}%
_{1}\otimes\mathcal{F}_{2})}  &  =\text{\ }b_{\nu_{e},s}^{+}(\mathbf{p})\text{
}\left\vert 0\right\rangle _{\mathcal{F}_{1}}\otimes\left\vert 0\right\rangle
_{\mathcal{F}_{2}},\\
b_{\nu_{\mu},s}^{+}(\mathbf{p})\left\vert 0\right\rangle _{C(\mathcal{F}%
_{1}\otimes\mathcal{F}_{2})}  &  =\text{ }b_{\nu_{\mu},s}^{+}(\mathbf{p}%
)\text{ }\left\vert 0\right\rangle _{\mathcal{F}_{1}}\otimes\text{\ }%
\left\vert 0\right\rangle _{\mathcal{F}_{2}}.%
\end{align}

Now, it is possible to define the number of rotated flavor particles as the
operator%
\begin{equation}
\varrho_{\theta}=\sum_{\mathbf{p}}\sum_{s=\pm1}(b_{\nu_{e},s}^{+}%
(\mathbf{p})b_{\nu_{e},s}(\mathbf{p})-b_{\nu_{\mu},s}^{+}(\mathbf{p}%
)b_{\nu_{\mu},s}(\mathbf{p})),
\end{equation}
which has eigenvectors and eigenvalues
\begin{align}
\varrho_{\theta}\text{ }b_{\nu_{e},s}^{+}(\mathbf{p})\left\vert 0\right\rangle
_{C(\mathcal{F}_{1}\otimes\mathcal{F}_{2})}  &  =b_{\nu_{e},s}^{+}%
(\mathbf{p})\left\vert 0\right\rangle _{C(\mathcal{F}_{1}\otimes
\mathcal{F}_{2})},\\
\varrho_{\theta}\text{ }b_{\nu_{\mu},s}^{+}(\mathbf{p})\left\vert
0\right\rangle _{C(\mathcal{F}_{1}\otimes\mathcal{F}_{2})}  &  =-b_{\nu_{\mu
},s}^{+}(\mathbf{p})\left\vert 0\right\rangle _{C(\mathcal{F}_{1}%
\otimes\mathcal{F}_{2})}.%
\end{align}

Since, the mentioned states are eigenfunctions of a physical observable (the
Hermitian operator $\varrho_{\theta}),$ the result of the measurement of the
rotated flavor eigenvalue should lead to the contraction of the wave-packet to
one of the eigenstates of $\varrho_{\theta}$ . Therefore, the probability of
the measurement will be the square of the amplitude defined by the scalar
product of those eigenstates and the eigenstate of the physical
quantity being measured.

It can be remarked that a similar transformation can be also implemented for
the antiparticle annihilation and creation operators $d_{\nu,s}(\mathbf{p})$
and $,d_{\nu,s}^{+}(\mathbf{p}).$

\subsection{Neutrino oscillations description: $\nu_{e}\rightarrow\nu_{e}$}

Let assume that an $electron$ $neutrino$ with helicity $s=-1$ had been
\ created over the vacuum at time equal to zero at the state
\begin{align}
\left\vert \phi_{\nu_{e},-1}(0)\right\rangle  &  =b_{\nu_{e},-1}%
^{+}(\mathbf{p})\text{ }\left\vert 0\right\rangle _{C(\mathcal{F}_{1}%
\otimes\mathcal{F}_{2})}\nonumber\\
&  =\left(  \cos(\theta)b_{1,s}^{+}\text{ }(\mathbf{p})+\sin(\theta
)b_{2,s}^{+}(\mathbf{p})\right)  \left\vert 0\right\rangle _{C(\mathcal{F}%
_{1}\otimes\mathcal{F}_{2})}.
\end{align}

Now, consider the evolution of the same $electron$ $neutrino$ density after a
time $t$. Then, acting with the evolution operator over the created state at
zero time, gives for the state at time $t$, the evolved state
\begin{align}
\left\vert \phi_{\nu_{e},-1}(t,\mathbf{p})\right\rangle  &  =U(t)\text{
}\left\vert \phi_{\nu_{e},-1}(0,\mathbf{p})\right\rangle \nonumber\\
&  =\exp(-i\text{ }H\text{ }t)\text{ }\left\vert \phi_{\nu_{e},-1}%
(0,\mathbf{p})\right\rangle \nonumber\\
&  =\left(  \exp(-i\epsilon_{1}(\mathbf{p})t)\cos(\theta)a_{1,-1}^{+}\text{
}(\mathbf{p})+\exp(-i\epsilon_{2}(\mathbf{p})t)\sin(\theta)a_{2,-1}%
^{+}(\mathbf{p})\right)  \text{ }\left\vert 0\right\rangle _{C(\mathcal{F}%
_{1}\otimes\mathcal{F}_{2})}.
\end{align}
where $\epsilon_{1}$ and $\epsilon_{2}$ are the energies of the mass eigenvalue neutrinos.

We can no examine  the projection amplitude of the above evolved state over the
$electron$ $neutrino$ state. Then, it is needed to evaluate the scalar product%
\begin{align}
_{C(\mathcal{F}_{1}\otimes\mathcal{F}_{2})}\left\langle 0\right\vert
b_{\nu_{e},-1}(\mathbf{p}^{\prime})\left\vert \phi_{\nu_{e},-1}(t,\mathbf{p}%
)\right\rangle  &  =_{C(\mathcal{F}_{1}\otimes\mathcal{F}_{2})}\left\langle
0\right\vert \left(  \cos(\theta)a_{1,-1}\text{ }(\mathbf{p})+\sin
(\theta)a_{2,-1}(\mathbf{p})\right)  \times\nonumber\\
&  \left(  \exp(-i\epsilon_{1}(\mathbf{p})t)\cos(\theta)a_{1,s}^{+}\text{
}(\mathbf{p})+\exp(-i\epsilon_{2}(\mathbf{p})t)\sin(\theta)a_{2,s}%
^{+}(\mathbf{p})\right)  \left\vert 0\right\rangle _{C(\mathcal{F}_{1}%
\otimes\mathcal{F}_{2})}\nonumber\\
&  =\cos(\theta)^{2}\exp(-i\epsilon_{1}(\mathbf{p})t)+\sin(\theta)^{2}%
\exp(-i\epsilon_{2}(\mathbf{p})t).
\end{align}

Therefore, the probability for the detection of the $electron$ $neutrino$ mode
at any time instant after its creation at zero time, becomes
\begin{align}
P_{\nu_{e}\rightarrow\nu_{e}}(t)  &  =\left\vert _{C(\mathcal{F}_{1}%
\otimes\mathcal{F}_{2})}\left\langle 0\right\vert b_{\nu_{e},-1}%
(\mathbf{p}^{\prime})\left\vert \phi_{\nu_{e},-1}(t,\mathbf{p})\right\rangle
\right\vert ^{2}\nonumber\\
&  =\left\vert \cos(\theta)^{2}\exp(-i\epsilon_{1}(\mathbf{p})t)+\sin
(\theta)^{2}\exp(-i\epsilon_{2}(\mathbf{p})t)\right\vert ^{2}\nonumber\\
&  =(\cos(\theta)^{2})^{2}+(\sin(\theta)^{2})^{2}+\nonumber\\
&  \cos(\theta)^{2}\sin(\theta)^{2}(\exp(-i\epsilon_{1}(\mathbf{p}%
)t+i\epsilon_{2}(\mathbf{p})t)+\nonumber\\
&  \exp(i\epsilon_{1}(\mathbf{p})t-i\epsilon_{2}(\mathbf{p})t))\nonumber\\
&  =(\cos(\theta)^{2})^{2}+(\sin(\theta)^{2})^{2}+\nonumber\\
&  2\cos(\theta)^{2}\sin(\theta)^{2}\cos\left(  (\epsilon_{1}(\mathbf{p}%
)-\epsilon_{2}(\mathbf{p}))t\right) \nonumber\\
&  =1-2\cos(\theta)^{2}\sin(\theta)^{2}\left(  1-\cos\left(  (\epsilon
_{1}(\mathbf{p})-\epsilon_{2}(\mathbf{p}))t\right)  \right) \nonumber\\
&  =1-\frac{\sin(2\theta)^{2}}{2}\left(  1-\cos\left(  (\epsilon
_{1}(\mathbf{p})-\epsilon_{2}(\mathbf{p}))t\right)  \right)  .
\end{align}

Let us consider now the relativistic approximation
\begin{equation}
|\text{ }\mathbf{p}\text{ }|\text{ }\gg m_{1},m_{2},
\end{equation}
which allows to derive the following  relation
\begin{align}
\epsilon_{2}(\mathbf{p})-\epsilon_{1}(\mathbf{p})  &  =\sqrt{m_{2}%
^{2}+\mathbf{p}^{2}}-\sqrt{m_{1}^{2}+\mathbf{p}^{2}}\nonumber\\
&  =|\mathbf{p|(}\sqrt{1+\frac{m_{2}^{2}}{\mathbf{p}^{2}}}-\sqrt{1+\frac
{m_{1}^{2}}{\mathbf{p}^{2}}})\nonumber\\
&  =\frac{1}{2|\mathbf{p|}}\mathbf{(}m_{2}^{2}-m_{1}^{2})+...
\end{align}

Then,  when the particles are ultra-relativistic, the
propagation time for traveling along a distance $R$ is given as
\begin{equation}
t=\frac{R}{c}=R,
\end{equation}
thanks to the natural units $c=1$ being used. Henceforth, the probability
formula for the transition between an $electron$ $neutrino$ state into another
$electron$ $neutrino$ state as a function of the measurement distance $R$ gets
the form
\begin{align}
P_{\nu_{e}\rightarrow\nu_{e}}(t)  &  =\left\vert _{C(\mathcal{F}_{1}%
\otimes\mathcal{F}_{2})}\left\langle 0\right\vert b_{\nu_{e},-1}%
(\mathbf{p}^{\prime})\left\vert \phi_{\nu_{e},-1}(t,\mathbf{p})\right\rangle
\right\vert ^{2}\nonumber\\
&  =1-\frac{\sin(2\theta)^{2}}{2}\left(  1-\cos\left(  -\frac{1}%
{2|\mathbf{p|}}\mathbf{(}m_{2}^{2}-m_{1}^{2})R\right)  \right) \nonumber\\
&  =1-\frac{\sin(2\theta)^{2}}{2}\left(  1-\cos\left(  -2\pi\frac{R}%
{L}\right)  \right)  ,\nonumber\\
L  &  =\frac{4\pi|\mathbf{p|}}{m_{2}^{2}-m_{1}^{2}}.
\end{align}
which reproduces the usual formula for the neutrino oscillations in terms of
the oscillation and observation distances $L$ $\ $and $R$ , the momentum
$|\mathbf{p|}$ and the neutrino masses $\ m_{1},m_{2}$.

In a similar way it can be evaluated the probability of measuring a muon
neutrino in the same state resulting form creating an electron neutrino at
zero time. The result is
\begin{align}
P_{\nu_{e}\rightarrow\nu_{\mu}}(t)  &  =\left\vert _{C(\mathcal{F}_{1}%
\otimes\mathcal{F}_{2})}\left\langle 0\right\vert b_{\nu_{\mu},-1}%
(\mathbf{p}^{\prime})\left\vert \phi_{\nu_{e},-1}(t,\mathbf{p})\right\rangle
\right\vert ^{2}\nonumber\\
&  =\frac{\sin(2\theta)^{2}}{2}\left(  1-\cos\left(  -2\pi\frac{R}{L}\right)
\right) \nonumber\\
&  =1-P_{\nu_{e}\rightarrow\nu_{e}}(t).
\end{align}

Therefore, the discussion presented  indicates that QFT or second
quantization is able explain the interference between distinguishable
particles which neutrino oscillation experiments show to exist. For this to
happens the Hilbert space should be considered as the linear completion of the
direct product of all the Fock spaces associated to each of the
indistinguishable particles included in the physical system. But, the use of
this space of states is naturally suggested if we do not assume the presence
of any superselection rule to restrict the superposition principle of the theory.

\section{Summary}

The question about why the neutrino oscillations effects curiously seems to
indicate the possibility of interference between distinguishable particles is
investigated. The compatibility of this effect with the quantum theory
defined by a QFT like the SM, is shown after a proper interpretation of the
space of states in the considered quantum field theory. The analysis was
suggested by the measured neutrino oscillations. A model involving only two
distinguishable types (flavors) of particles is formulated, in order to
consider the study within a simple framework. It consists of two types of spin
one-half  relativistic particles satisfying the Dirac equation.

The formulation of the space of the states of the system is examined.  It is
argued that in order to describe interference between two fermion flavors in
analogy with the neutrino oscillation measurements, the space of states should
not correspond to the direct product of the two Fock spaces associated to the
two fermion particles. It is underlined that to employ the direct product as
the space of states leads to accept the unphysical character of the addition
of two field operators, each one creating a particle in the two separate Fock
spaces. The reason for this is that the states generated by the action of such
additions do not pertain to the direct product of the two Fock spaces. To
exclude these states from the space of physical states becomes equivalent to
establish the validity of a superselection rule.

Therefore, we passed to consider as the space of the states the linear
completion of the direct product of the two Fock states. That is, the linear
combination with arbitrary coefficients of the external products of the states
in each of the two Fock spaces. It is then a verified that this space allows
to define the addition of field operators acting in each separate Fock space,
as generating new physical states. Further, rotated flavor creation and
annihilation operators are defined by linear combinations of the creation and
annihilation operators for propagating modes solving the Dirac equation. This
procedure allows to argue that states generated by these rotated flavor fields
at a given time, are \textit{physical} linear combinations of the propagating
particles modes. In addition, it follows that under a measurement in such a
propagating states (defining after it a contracted flavor rotated state) the
probability of the measurements oscillates. The evaluated expression for the
transition probabilities exactly coincide with ones describing the observed
neutrino oscillations.

The analysis clarifies that the usual representation of QFT allows to
consistently predict quantum interference between two distinguishable
particles, explaining in this way how the quantum theory which is a QFT, can
explain the interference between different particles, at variance with what is
sometimes had been considered in QM. It also suggests the validity of what we
consider a relevant property of QFT generalization of the Quantum Mechanics:
the physical possibility of observing interference between many kinds of
distinguishable particles. This indication comes from the fact that the wider
superposition principle in QFT (with respect to QM)\ allows to add states
having components in different Fock spaces (distinguishable particles) and
then, interference effects should be expected to be observable in seemingly
many processes. \ These issues will be further explored in \ elsewhere.

\section*{Acknowledgements}

A.C. is grateful to the Department of Physics of the "Division de Ciencias e
Ingenierias de la Universidad de Guanajuato by the support to this research
during a short visit to this Center in Nov. 2109. He will also acknowledge the
additional support received from the Proyecto Nacional de Ciencias B\'{a}sics
(PNCB, CITMA, Cuba) and from the Network N-09 of the Office of External
Activities of the ICTP.

\end{document}